\documentstyle[12pt,amssymbols]{article}

\def\ie{\hbox{\it i.e.}}

\def\np#1#2#3{           {\it Nucl. Phys. }{\bf #1}, #2 (19#3)}
\def\pl#1#2#3{           {\it Phys. Lett. }{\bf #1}, #2 (19#3)}
\def\pr#1#2#3{           {\it Phys. Rev. }{\bf #1}, #2 (19#3)}

\def\prl#1#2#3{          {\it Phys. Rev. Lett. }{\bf #1}, #2 (19#3)}

\def\sjn#1#2#3{           {\it Sov. J. Nucl. Phys. }{#1}, #2 (19#3)}

\def\zp#1#2#3{           {\it Zeit. fur Physik }{\bf #1}, #2 (19#3)}

\def\abs#1{\left| #1\right|}
\def\etal{\hbox{\it et al.}}
\input{epsf}
\begin{document}
\begin{titlepage}
\begin{center}
\today     \hfill    LBL-36374 \\

\vskip .5in

{\large \bf Current Issues in Perturbative QCD\footnote{Invited talk
presented at
the 1994 Meeting of The Division of Particles and Fields of the American
Physical Society, Albuquerque NM, August 1-6, 1994}}
\footnote{This work was supported by the Director, Office of Energy
Research, Office of High Energy and Nuclear Physics, Division of High
Energy Physics of the U.S. Department of Energy under Contract
DE-AC03-76SF00098.}
\vskip .50in

\vskip .5in
Ian Hinchliffe \\
{\em Theoretical Physics Group\\
    Lawrence Berkeley Laboratory\\
      University of California\\
    Berkeley, California 94720}
\end{center}

\vskip .5in

\begin{abstract}
This review talk discusses some issues of active research in
perturbative QCD. Among the topics discussed are, heavy flavor and prompt
photon production in
hadron-hadron collisions, ``small $x$'' phenomena and the current status of
$\alpha_s$.
\end{abstract}
\end{titlepage}

\renewcommand{\thepage}{\roman{page}}
\setcounter{page}{2}
\mbox{ }

\vskip 1in

\begin{center}
{\bf Disclaimer}
\end{center}

\vskip .2in

\begin{scriptsize}
\begin{quotation}
This document was prepared as an account of work sponsored by the United
States Government. While this document is believed to contain correct
 information, neither the United States Government nor any agency
thereof, nor The Regents of the University of California, nor any of their
employees, makes any warranty, express or implied, or assumes any legal
liability or responsibility for the accuracy, completeness, or usefulness
of any information, apparatus, product, or process disclosed, or represents
that its use would not infringe privately owned rights.  Reference herein
to any specific commercial products process, or service by its trade name,
trademark, manufacturer, or otherwise, does not necessarily constitute or
imply its endorsement, recommendation, or favoring by the United States
Government or any agency thereof, or The Regents of the University of
California.  The views and opinions of authors expressed herein do not
necessarily state or reflect those of the United States Government or any
agency thereof or The Regents of the University of California and shall
not be used for advertising or product endorsement purposes.
\end{quotation}
\end{scriptsize}

\vskip 2in

\begin{center}
\begin{small}
{\it Lawrence Berkeley Laboratory is an equal opportunity employer.}
\end{small}
\end{center}

\newpage
\renewcommand{\thepage}{\arabic{page}}
\setcounter{page}{1}

\section{The current value of $\alpha_s$}

I will present a brief update of the value of $\alpha_s(M_Z)$. For a review
of
the results prior to this meeting see the article in QCD in the 1994
edition of
the Review of
Particle Properties \cite{rpp}.
The methodology adopted there will be followed here. Results from
experiments
using similar methods that have common systematic errors are first
combined.
 These results are then
extrapolated up to the $Z$ mass using the renormalization group.
An average of these values
is then made to give the final result which is quoted as a value for
$\alpha_s(M_Z)$. The new results will now be discussed.

\subsection{Lattice Gauge Theory.}

Lattice gauge theory  calculations can be used to  calculated the energy
levels
of a   $Q\overline{Q}$  system and  then  extract   $\alpha_s$. The  FNAL
group
  \cite{Khadra93} uses the  splitting between the
1S and  1P in the charmonium system
($m_{h_c}-(3m_{\psi}+m_{\eta_c})/4=456.6\pm     0.4$ MeV).
to determine $\alpha_s$. The result quoted is
$\alpha_s(M_Z)=0.108\pm 0.006$.  The  splitting is almost
independent   of the  charm  quark  mass and  is  therefore  dependent
only on
$\alpha_s$.
The   calculation  does not  rely on    perturbation  theory or on
non-relativistic approximation. The  main errors are systematic associated
with
the  finite  lattice  spacing  ($a$), the matching to the perturbatively
defined
$\alpha_s$, and  quenched   approximation  used in the
calculation.  The  extrapolation to  zero lattice  spacing  produces a
shift in
$\Lambda$ of order 5\% and is therefore quite small. The quenched
approximation
is more   serious. No  light  quarks are  allowed  to  propagate and  hence
the
extracted   value of   $\Lambda$   corresponds to  the  case of  zero
flavors.
$\alpha_s(M)$ is  evolved down from  the scale ($\sim 2.3$  GeV) of the
lattice
used to the scale  of momentum  transfers appropriate to  the charmonium
system
($\sim  700$ MeV).  The  resulting coupling  is then  evolved back  up with
the
correct number of quark flavors.
Perturbative running of $\alpha_s(M)$ has to be used at small $M$.

A recent calculation \cite{luscher} using using the strength of the force
between two heavy quarks computed in the quenched approximation obtains a
value of
$\alpha_s$ that is consistent with this result.

Calculations
based on the $\Upsilon$ spectrum using non relativistic lattice theory give
$\alpha_s(M_Z)=0.115\pm 0.003$   \cite{lepage}. This result includes
relativistic corrections up to order $m_b(v/c)^4$.
This recent result does not rely on the quenched approximation.
Calculations
are performed with two massless flavors. Combining this with the result in
quenched approximation  enables the result to be extracted for the
physical case of three ($u,d$ and $s$) light quarks.
It is gratifying that this result is within the error quoted on the
quenched
calculations \cite{davies}.
 Averaging the lattice results then yields $\alpha_s(M_Z)=0.113\pm 0.003$
\begin{figure}
\epsfbox{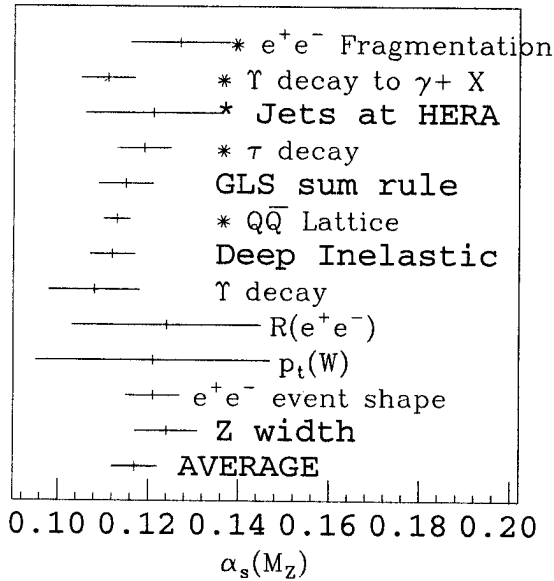}
\caption[]{The values of $\alpha_s(M_Z)$ determined by various methods. The
symbol $*$ denotes a result that has been updated from that in Ref
\cite{rpp}.
}
\label{fig1}
\end{figure}

\subsection{Jet counting}
A recent result from CLEO \cite{cleo} measuring jet multiplicities at
 in $e^+e^-$
annihilation at $\sqrt{s}=10 $ GeV, \ie\  below the $b\overline{b}$
threshold
gives  a result of $\alpha_s(M_Z)=0.113\pm 0.006$. As with all measurements
of
this type, the dominant errors are systematic and arise from ambiguities in
the
scale at which $\alpha_s$ is evaluated and from the algorithms used to
define a
jet. This result is consistent with that from higher energies, in
particular
those from LEP\cite{lepshape} and SLC \cite{sldshape}.

Data at $\sqrt{s}=29,$ $58$ and $91$
GeV have been fit with the same set of Monte-Carlo (fragmentation)
parameters.
A consistent fit is obtained providing direct evidence for the running of
$\alpha_s(Q)$. \cite{ronan}

The H1 collaboration working at HERA \cite{h1} has determined $\alpha_s$
from a
fit to the $2+1$ jet rate \cite{inglemann}.
At lowest order in QCD the final state in deep inelastic scattering
contains
$1+1$ jets, one from the proton beam fragment and one from the quark that
is
struck by the electron. At next order, the struck quark can radiate another
gluons giving rise to the $2+1$ jet final state. The
 determination involves data
over a large $Q^2$ range and hence there is some correlation between the
value of $\alpha_s$ and the structure functions that enter the computation
of
the event rate.
 The value quoted is $\alpha_s(M_Z)=0.121\pm 0.015$.

\subsection{Upsilon decay.}

The Cleo group\cite{cleogamma}
has determined $\alpha_s$ from a measurement of the ratio of
Upsilon decay rates $\frac{\Upsilon\to \gamma +hadrons}{\Upsilon \to
hadrons}$.
In lowest order QCD this is given by $\frac{\Upsilon\to \gamma gg
}{\Upsilon
 \to ggg}$. They quote $\alpha_s(M_Z)=0.111\pm 0.006$. There is
 non-perturbative contribution to this
final state from the fragmentation of a
 gluon jet into a photon; this will introduce additional systematic errors
into
 the result.

\subsection{Scaling violations in Fragmentation functions.}

The  probability for a  quark produced  at scale  $Q$ (for  example in
$e^+e^-$
annihilation at  $\sqrt{s}=Q$) and energy $E$ to  decay into a hadron of
energy
$zE$ is parameterized by a fragmentation function $d(z,Q)$. Just as in the
case
of the structure  functions, the $Q$ dependence of  this fragmentation
function
is given by perturbative QCD and  depends only on $\alpha_s$. The QCD
evolution
of this  fragmentation function also  involves the  fragmentation function
of a
gluon ($g(z,Q)$).  Hence in order to  determine $\alpha_s$  both $d(z,Q_0)$
and
$g(z,Q_0)$   must be   determined at  some  reference  point  $Q_0$.  The
ALEPH
collaboration\cite{alephfrag}  uses  three jet events from  the decay of a
$Z$.
Two of the jets are tagged to be from  $b-$quarks using the vertex detector
and
the finite  $b-$quark lifetime. The  third is then known  to be due to a
gluon.
This method also  determines the  fragmentation functions  for charm and
bottom
quarks which  do not have  the same  form at $Q_0$  as the light  quarks.
It is
worth  recalling  that  whereas  higher  twist  corrections  in deep
inelastic
scattering  are of  order  $1/Q^2$,  here they  can be  order $1/Q$.  These
are
parameterized in the  ALEPH fit by replacing $z$ by  $z+c(z)/Q$. ALEPH
quotes a
value of   $\alpha_s(M_Z)=0.127\pm  0.011$. The  DELPHI  collaboration,
using a
different method quotes  $\alpha_s(M_Z)=0.118\pm 0.005$ \cite{delphifrag}.
This
result  does not use  the  independent  measurements of  heavy  quark and
gluon
fragmentation  functions but rather  fits to a  Monte-Carlo. Its error
could be
underestimated.

\subsection{Hadronic Width of the Tau Lepton}

The  hadronic  width (or  branching  ratio) of  the tau  lepton can  be
used to
determine    $\alpha_s$   \cite{braaten}.  In the  decay   $\tau\to
\nu_{\tau}
+hadrons$,  the  decay  rate,  $R(M)$, can  be  measured as  a  function of
the
invariant mass $M$ of the hadronic system. The inclusive hadronic width is
then
obtained by integrating  over $M$, {\it viz.}  $\Gamma=\int dM R(M)$. There
are
non-perturbative (higher twist)  contributions that can be calculated using
QCD
sum rules \cite{sumrule}. Alternatively the data can be used to determine
these
quantities,   which have   different $M$   dependence, from  the data  by
using
$\gamma(n)=\int dM M^n  R(M)$. The values obtained  in this are consistent
with
the estimates from  the sum rules.   \cite{cleotau}\cite{alephtau}  There
is an
new result from CLEO \cite{cleotau}  that gives $\alpha_s(M_Z)=0.114\pm
0.003$,
a  value  somewhat  below  the old  world   average. A  new  result  from
ALEPH
\cite{alephtaunew}  of   $\alpha_s(M_Z)=0.124\pm 0.003$ is  larger than the
old
world average. The difference between  these results is due to different
values
of  the   branching  ratios  $R_e$  and   $R_{\mu}$   measured  for
$\tau\to e
\overline{\nu}\nu\overline{\nu}$                             and
$\tau\to
\mu\overline{\nu}\nu\overline{\nu} $.  The hadronic width is then inferred
from
these via    $R_h=1-R_e-R_{\mu}$ as this  results in a smaller  error than
that
gotten by using $\Gamma$ directly.

\subsection{Average value of $\alpha_s(M_Z)$}

After taking into account this new data, the average of
$\alpha_s(M_Z)=0.117$
quoted in RPP 1994 is left unchanged. If we assume that the systematic
errors
associated with the different methods are uncorrelated, then we obtain an
error of $\pm 0.002$. In view of the fact that most of the dominant errors
are
theoretical, involving such things as estimates of non-perturbative
corrections
and the choice of scale $\mu$ where $\alpha_s(\mu)$ is evaluated for the
process in question, it is more reasonable to quote $\alpha_s(M_Z)=0.117\pm
0.
005$ as the ``world average''.

\section{Heavy Quark Production in Hadron Collisions}

New results are available from the CDF\cite{cdfb} and D0\cite{d0b}
 collaborations on the production
rate
of bottom quarks in $p\overline{p}$ collisions. Several methods are used.
The
least subject to ambiguities involves the use of
fully reconstructed decays of a B
meson. In this case in order to get from the observed B meson rate to that
of
$b-$quarks, the fragmentation function of a $b-$quark needs to be known.
This
is well constrained from data at LEP so this method of measuring the
$b-$quark
production rate should be quite reliable. However there are rather few
fully
reconstructed events and hence this method is limited and does not
permit measurement over a large range of transverse momenta.
\begin{figure}
\epsfbox{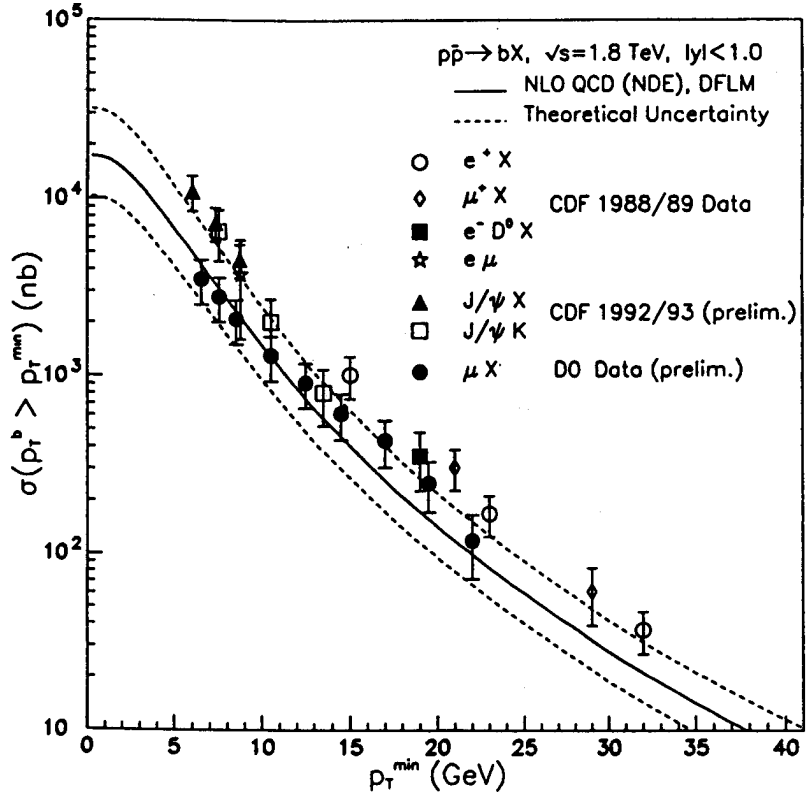}
\caption[]{{The inclusive cross section for the production of $b$-quarks in
$p\overline{p}$ collisions at $\sqrt{s}=1.8$ TeV. The produced quark is
required to have transverse momentum greater than $p_T^{min}$ and the rate
is
shown as a function of $p_T^{min}$. See text for discussion.}}
\label{figbot}
\end{figure}

The next method involves the use of
inclusive $\Psi$ production. This method can only be used if a vertex
system is available to disentangle the $\Psi$'s that come from $b$-decay
from
those produced directly. Here the systematic errors are a larger since one
needs a model of the $b-$quark fragmentation and of the subsequent
$b-$meson
and $b-$baryon decays to $\Psi X$.

 Finally, there is the method with the largest statistical sample and hence
the
greatest range in  transverse  momentum. Here one searches  for jets which
have
muons or electrons associated with  them. The leptons are required to have
some
transverse momentum of order 1 GeV or greater with respect to the jet
direction
($p_a$). Most of these  leptons then arise from  bottom decay: there is a
small
contribution from charm decay, the relative fraction being a function of
$p_a$.
A model of the  lepton spectrum from  charm and bottom  quarks is needed
before
the $b$-quark cross section can be extracted.

The measured rates from the different  methods are shown in figure
\ref{figbot}
which shows the  cross-section for the production  of a $b$-quark of
transverse
momentum  greater that  $p_T^{min}$. It can be  seen from this  figure that
the
rates measured by  the $D0$  collaboration which uses only  the last method
are
systematically lower than those of the  CDF collaboration which uses all of
the
methods for $b$-quark transverse  momenta of less than 15 GeV. Above this
value
the  experiments are  consistent with  each other.  Note that in  the
region of
disagreement CDF  is able to use what  should be the most  reliable method.
The
figure also shows the  theoretical expectation for  this rate \cite{nde}.
It is
rather uncertain since  the QCD predictions are not  stable with respect to
the
choice of the scale $\mu$ at which the parton distributions and
$\alpha_s(\mu)$
is evaluated  in the  expression for the  production rate. By  lowering
this to
$\mu=\sqrt{m_b^2+p_t^2}/4$  consistency with the  CDF data can be achieved.
The
D0 data can be accommodated by using  larger and {\it a priori} more
reasonable
value of $\mu$. The cross-section now  reported by CDF is lower than the
values
that were obtained  from observation  of $\Psi$ production  and the
assumption,
now known to  be wrong, that  almost all  $\psi$'s at large  transverse
momenta
arise from $b$-quark decay.

The top cross-section  quoted by CDF\cite{cdftop}  is somewhat larger than
that
predicted by QCD for a mass of 170 GeV\cite{laenen}, the value given by the
CDF
fit. Since D0 has not yet  confirmed this  rate\cite{d0top}, it is
premature to
claim that  there is a  problem with  QCD or that  physics  beyond the
standard
model has been discovered.

\section{Production of $\Psi$ and $\Upsilon$ in $p\overline{p}$
collisions.}

At low transverse momentum the production of $\psi$'s in $p\overline{p}$
collisions is expected to proceed dominantly via the production of $\chi$
states
followed by their decay \ie \
$g+g\to\chi\to \psi +X$\cite{stirling}. The analogous process at large
transverse momentum is $gg\to g\chi$. This process generates a cross
section
that falls off at large transverse momentum much faster than, say, the jet
rate.
This observation led to an assumption that $\psi$'s produced at large
transverse
momentum came almost exclusively from $b$-quark decay. A measurement of the
rate
for $\psi$ production could then be used to infer the $b$-quark rate. This
assumption is now known to be false. By detecting whether or not the
$\psi$'s
come from the primary event vertex, CDF is now able to test this
assumption.
The fraction of $\psi$'s produced directly is almost independent of
transverse momentum and the rate of direct $\Psi$ production
at large $p_t$ is larger than had been
expected.

The dominant production mechanism of $\psi$'s at large transverse momentum
is
now believed to be the fragmentation of light quark and gluon jets into
$\chi$'s that then decay to $\psi$ \cite{fleming}. CDF now has data on
$\psi$,
 $\psi$, $\psi^{\prime}$, $\Upsilon$, $\Upsilon^{\prime}$ and
 $\Upsilon^{\prime\prime}$ \cite{vaia}. While
 the rate for $\psi$ production is in
 agreement with expectations, given the inherent theoretical uncertainties,
the
 rate for $\psi^{\prime}$ is approximately a factor of 20 above the
theoretical
 expectation \cite{fleming1}. The calculation does not include the
possibility
 of $\psi^{\prime}$ production from the decay of 2P states. These states
are
 above the $D\overline{D}$ threshold, however a branching ratio of a few
 percent to $\psi^{\prime}$  could be enough to explain the deficit. A
recent
paper investigates this possibility quantitatively \cite{roy}

The predicted rates for $\Upsilon$ production should have less
uncertainties
due to the larger value of the $\Upsilon$ mass. Preliminary data from CDF
indicate that the agreement with theoretical expectations is poor
\cite{vaia}.

\section{Prompt Photon Production.}

The production of photons in $p\overline{p}$ proceeds, at lowest order in
QCD,
via the parton process $qg\to \gamma q$. The process provides a direct
probe of
the gluon distribution and can be measured more reliably than the jet
cross-section whose value depends on a jet definition and upon measurements
of
both hadronic and electromagnetic energy. The produced photon, provided
that is
is produced at large transverse momentum, is well isolated from other
produced
particles.  At higher orders in $\alpha_s$, the
situation changes. Processes such as $qq \to qq\gamma$ start to contribute.
This process is largest when the photon is collinear with one of the
outgoing
quarks. Since experiments cannot easily measure
photons within jets, they search for isolated photons defined by having
less
than some amount $\epsilon$ of other energy in a cone of radius $\Delta R$
in rapidity-azimuth
 space around the photon direction. The rate then depends on
$\epsilon$ and $\Delta R$; the selection criteria discriminate against the
bremsstrahlung component. The fragmentation of a quark into a photon is a
non-perturbative phenomena which must be modeled by a fragmentation
function
into which the collinear singularity is absorbed.

A theoretical prediction of the prompt photon rate then depends on, the
gluon and quark distributions, the fragmentation functions, and the scales
$\mu$ and $Q$
at which these functions and $\alpha_s(Q)$ are evaluated
(these scales need not be the
same).
The dependence on these scales is an indication of the uncertainties in the
theoretical predictions; if the process were calculated to all orders in
perturbation theory, the dependence on $\mu$ and $Q$ would, at least in
principle, disappear.
$\mu$ and $Q$ should be of order $p_t$, beyond that theory provides no
guidance. There is a longstanding problem in that at small values of
$x_{\perp}=2p_t/\sqrt{s}$, the data tend to be larger than the theoretical
predictions. At values of $x_{\perp}$ that are probed at the Fermilab
collider
a smaller value of $\mu$ results in a larger predicted cross-section.

There  have  been  several new   developments in  this  field.
Measurements of
structure functions at small $x$ at  HERA\cite{herarev} have indicated that
the
gluon distribution  is larger at small  values of $x$ than  used to be
assumed.
This change increases the predicted  rate at small $x_{\perp}$.  There has
been
a reassessment of the importance of  fragmentation \cite{vogelsang} and
finally
new data are available.

\begin{figure}
\epsfbox{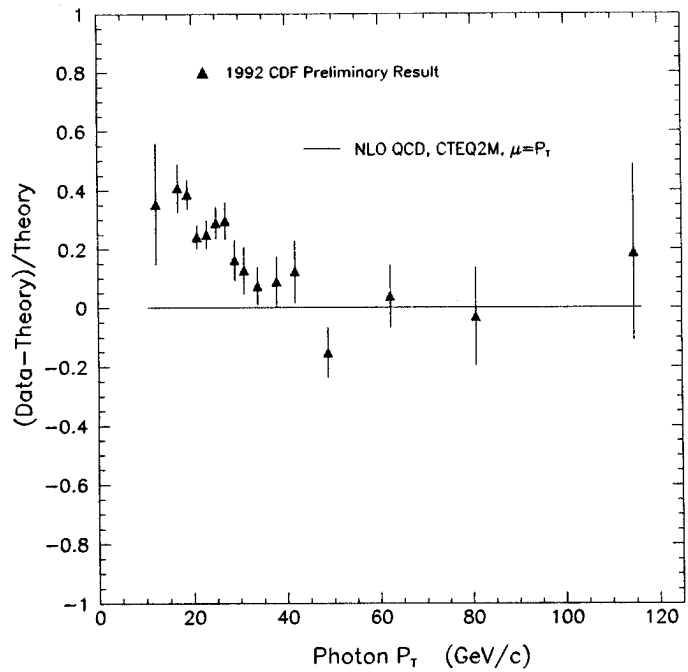}
\caption[]{A comparison of the CDF data on prompt photon production.
 Data refers
to the quantity $\frac{d\sigma}{dp_t d\eta}$ at $\eta=0$.
Theory refers to the calculation of  Ref
\cite{owens}
 using the CTEQ2M
structure functions
[36] and $\mu=p_t$.}
\label{phot1}
\end{figure}

\begin{figure}
\epsfbox{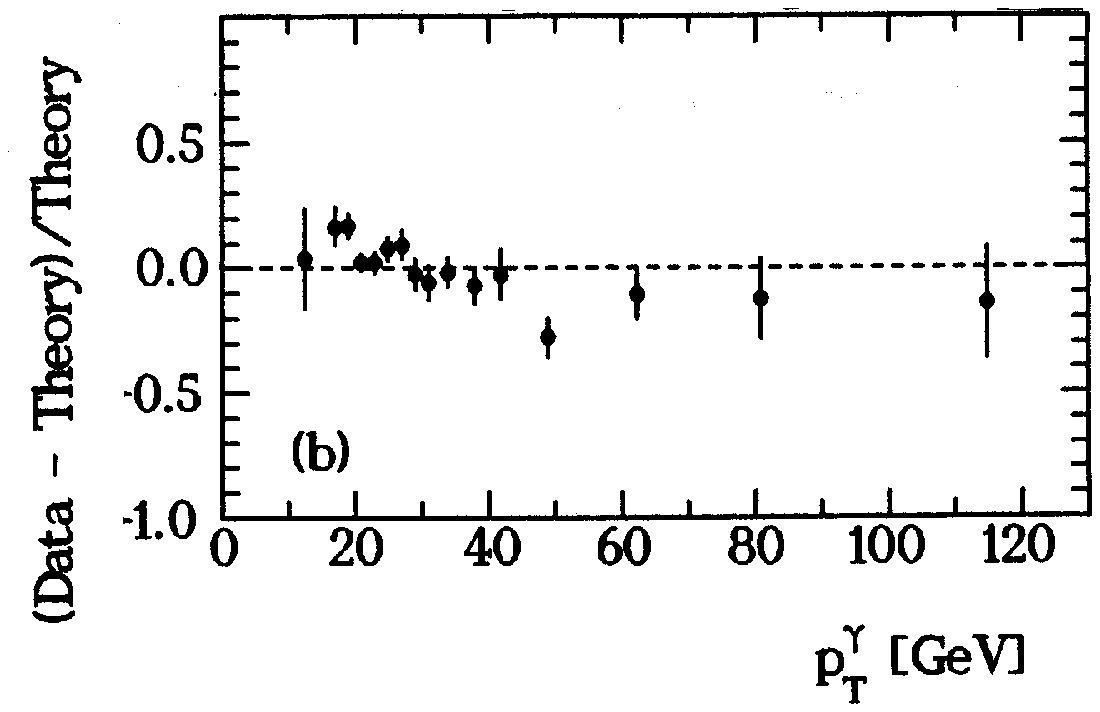}
\caption[]{As Figure \ref{phot1} except
 that the theory refers to the calculation of  Ref
\cite{vogelsang}
 using
the GRV  structure functions
[38]
 and $\mu=p_t/2$.}
\label{vogfig}
\end{figure}

Figure \ref{phot1} shows that data from the CDF collaboration
\cite{cdfgamma}.
 The data fall
very rapidly with increasing $p_t$ so, to facilitate comparison with
theory,
the data are shown relative to the calculation of ref \cite{owens}. The
tendency for the data to have a steeper dependence on $p_t$ than the theory
can
be seen in this figure. A reduction in the scale to $\mu=p_t/2$ brings the
data
into better agreement with the theory.
Figure \ref{vogfig} shows the same data
compared to the calculation of ref \cite{vogelsang}. Here the scale
$\mu=p_t/2
$ has been used along with the GRV structure functions \cite{grv}. The
predictions of $MRSD-^{\prime}$ structure functions\cite{mrs}
 are almost identical. This
theoretical result has a slightly larger fragmentation component. The
tendance
for the the data to have a steeper $p_t$ slope than the theory is still
apparent in this plot, although the agreement is quite good. A reduction in
the
scale $\mu$ to the, possibly unreasonable, value of $p_t/3$ improves the
agreement further.

\begin{figure}
\epsfbox{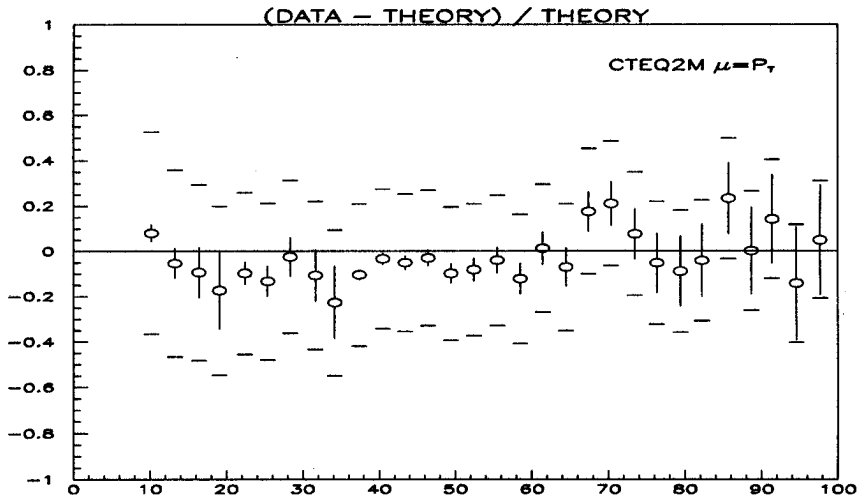}
\caption[]{As Figure \ref{phot1}
except that the experiment refers to the results of the D0 collaboration
\cite{d0gamma}
}
\label{d0gamma1}
\end{figure}

Preliminary results presented at this meeting from $D0$\cite{D0gamma} lie
somewhat below those of CDF and are therefore in better agreement with
theoretical estimates. Figure \ref{d0gamma1} shows a comparison with
theory.
Note
that that theory is the same as that used in figure \ref{phot1}.
The tendency for the data to have  a steeper $p_t$ slope than the theory is
not
evident in the D0 results although
the systematic errors are such that there is no
significant disagreement with CDF.

There is a preliminary measurement of  the di-photon rate from
CDF\cite{blair}.
This measurement  is important since,  at the LHC, one of  the decay
mechanisms
proposed to search for  the Higgs boson is its  decay to two
photons\cite{lhc}.
The  very  large   background is  expected  to  occur  from
$q\overline{q}\to
\gamma\gamma$ and $g g\to \gamma\gamma$. The rate observed by CDF is
consistent
with the expectation from calculations using these
mechanisms\cite{gampair}. We
can now have more confidence in the ability of LHC to see the Higgs signal.

\section{Small-$x$ and related phenomena}

QCD perturbation theory is an expansion powers of $\alpha_s(Q)$. Two
conditions
must be satisfied to have  a reliable prediction.  First, the scale $Q$
must be
large and second  the perturbation  series must contain no  large
coefficients.
The value of some  measured  dimensionless  quantity $P$  and be expressed
as a
power series.  $$P=A\alpha_s^n(Q)(1+b\alpha_s(Q)+\cdots)$$ If
$b\alpha_s(Q)\sim
1$, then  the  perturbation  series  useless.  A  physical  prediction can
be
recovered if  the large terms  can be isolated  order by order  in
perturbation
theory and the  terms summed up. Once  these pieces are  absorbed into $A$,
the
resulting series may be well behaved and a prediction possible.

The simplest example of a resummation of this type  is that of the
Altarelli-Parisi (DGLAP) equation which sums terms of the type
 $(\alpha_s \ln(Q^2))^n$
that arise in Deep Inelastic Scattering \cite{glap}. At
very small values of Bjorken-$x$, $b$ can contain terms of the
type $\log (1/x)$. These terms can be resummed using the BFKL equation
\cite{bfkl}. The result of this resummation is a structure function that
rises
very rapidly at small-x. While this behaviour is seen at
HERA\cite{herarev},
it cannot be used to distinguish between evolution expected from BFKL and
DGLAP\cite{ellis}.

The behaviour of the structure functions at very small $x$ is connected
with
attempts to calculate the total cross-section in perturbative QCD. The same
resummation that leads to BFKL is also responsible for the appearance in
perturbative QCD of the pomeron\cite{lipatov}. This connection has recently
been clarified\cite{mueller1}. I will now discuss some phenomena related to
the pomeron.

\subsection{Jets with Large Rapidity Separation}

Events are selected in $p\overline{p}$ collisions having a pair of jets
with
transverse momenta $p_1$ and $p_2$ and rapidities $\eta_1$ and $\eta_2$
with
azimuthal angle $\phi$ between then. At
lowest order in perturbative QCD, $p_1=p_2$,  $\cos(\pi-\phi)=1$ and the
rate is
given by
$$\frac{d\sigma}{dp_1d\eta_1d\eta_2}=\frac{1}{16\pi s}g(x_1,Q^2)g(x_2,Q^2)
\frac{d\sigma(gg\to gg)}{dt}$$
Here I have  assumed that only gluons  contribute. If  $\eta_1=-\eta_2=y$,
then
$x_1=x_2=\frac{2p_1}{\sqrt{s}}cosh{y/2}$.  If $y$  is very large, the
center of
mass energy of the  parton system  ($=x_1x_2s$) becomes  large and the
partonic
cross                 section      can     be                 approximated
 by
$$\frac{d\sigma}{dt}=\frac{9\pi\alpha_s^2}{2p_1^2 p_2^2}$$  If we now
integrate
over       $p_1$    and     $p_2$       greater    than    some     scale
$M$
$$\frac{d\sigma}{d\eta_1\eta_2}\sim
\frac{\alpha_s^2}{M^2}x_1g(x_1)x_2g(x_2)$$

At order $\alpha_s^3$ in perturbation  theory, several phenomena occur. A
third
jet is emitted and the  correlation in azimuth and  equality of $p_1$ and
$p_2$
is lost. More  important is that this  order $\alpha_s^3$  process modifies
the
result   for            $\frac{d\sigma}{dp_1d\eta_1d\eta_2}$  by  a
factor of
$(1+3\alpha_s\abs{\eta_1-\eta_2}/\pi)$  (for large  values of
$\eta_1-\eta_2$).
If    $\eta_1-\eta_2$  is  large  enough  this   factor can  be  so  large
that
perturbation  theory is  not reliable.  In this  case the  leading terms at
all
orders   in    perturbation  theory  can  be   resummed  to  give a
factor of
$exp(3\alpha_s\abs{\eta_1-eta_2}/\pi)$      \cite{mueller}. This  growth is
not
observable at the  Tevatron since it  is more than  compensated by the drop
off
caused by the falling structure  functions. (Note that $x_1$ and $x_2$
increase
as $y$  increases.). It  may be  observable at LHC   \cite{schmdt}. However
the
other effects should be observable. The rapidity region between the two
jets is
filled with many  mini-jets since there is no  penalty of $\alpha_s$ to pay
for
each emission. The  correlation in  $\phi$ between the two  trigger jets
should
show a rapid fall off as $y$ is increased. The D0 collaboration\cite{d0rap}
has
searched for  this effect by  selecting events  with two jets  one of which
has
$p_t> 20$GeV and  the other has $p_t>  50$ GeV. The $\phi$  correlation is
then
plotted   as a   function  of  the   rapidity    separation.  The  data
show a
decorrellation.  However  it is a much  slower fall  off than  predicted
and is
consistent with  that expected from a  fixed order  $\alpha_s^3$
calculation or
from showering  Monte-Carlos such as  HERWIG\cite{herwig}.  It is possible
that
the rather asymmetric trigger could be masking the effect in this case.

\subsection{Rapidity Gaps.}

Consider   the  production  of two  jets  at  large  rapidity   separation
in a
$p\overline{p}$   collision. At  lowest  order in QCD   perturbation theory
one
contribution  to this arises  from quark-quark  scattering via  gluon
exchange.
Before the scattering  each quark forms a color  singlet state with the
rest of
the quarks and gluons from its parent  (anti-)proton. After scattering,
this is
no  longer  the case  since  the  gluon  exchange   causes  color  charge
to be
transferred  between the  quarks. As  the parton  system  hadronizes into
jets,
color  must be   exchanged  between  the  outgoing  jets. This  color
exchange
manifests  itself as soft  (low  transverse)  momentum particles  that fill
the
rapidity  interval  between the  jets.  Contrast this  with the  situation
if a
colorless object  (such as a photon)  were exchanged. Now  the struck quark
and
the  remnant of  its parent  are still  in a  color  singlet and can
hadronize
without  communication with  the other quark.  There is no  necessity for
color
exchange and  hence no need  for particle  production in the  rapidity
interval
between the jets. Both  the CDF\cite{devlin} and D0\cite{d0gap}
  collaborations have searched for events
with rapidity  gaps. CDF uses the  charged particle  multiplicity while D0
uses
the energy flow as measured by the calorimeter.

\begin{figure}
\epsfbox{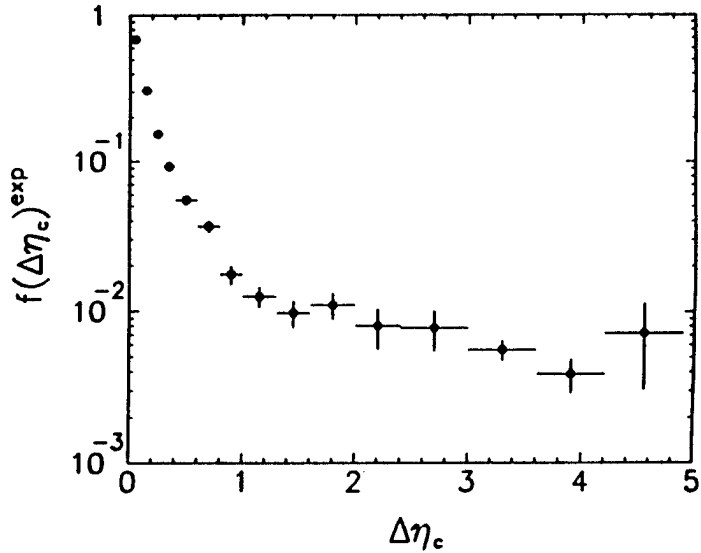}
\caption[]{The fraction of events with no particles in the rapidity
interval
between the two produced jets in a $p\overline{p}$ collision as a function
of
the rapidity interval. Data from the D0 collaboration
\cite{d0gap}.
}
\label{gap1}
\end{figure}

In the D0 case, events  with two jets each of  transverse energy of at
least 30
GeV are   selected. The  jets are   separated by  rapidity  $\eta$.  Events
are
determined  to have a gap  if  there are no calorimeter towers in the
region
between the two jets with an electromagnetic energy deposit of more than
200
MeV. Figure  \ref{gap1}  shows the fraction
($f$) of events that have such gaps as a function of the rapidity
separation of
the  jets. If all  of the  events are  due to  jet  production  involving
color
exchange, one expects that $f$ will  fall rapidly with increasing $\eta$.
While
this  behaviour is  observed  at small  $\eta$  there is  clear  evidence
for a
plateau in $f=f_0$ at  large values of $\eta$  indicating the presence of
color
singlet exchange.

\begin{figure}
\epsfbox{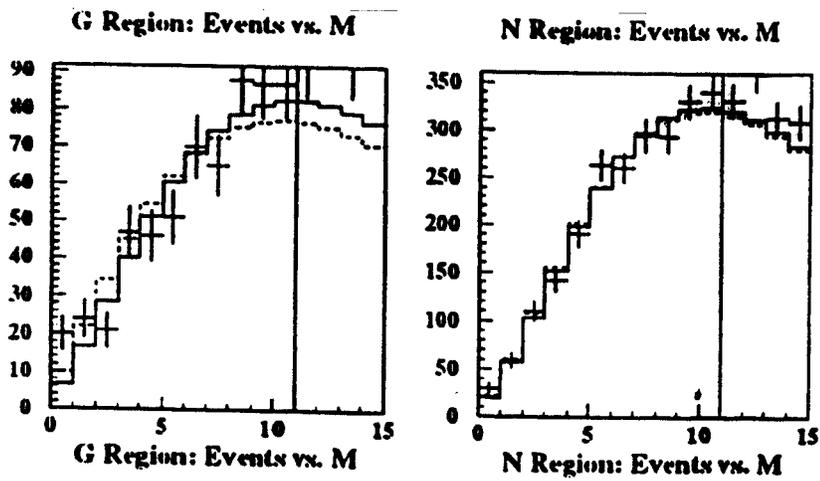}
\caption[]{The event rate plotted against particle multiplicity. Two jets
are
selected, separated in rapidity by 2.8 units of rapidity. The G (N)
region is defined as the interval between the jets (between each jet and
the
end of the physical region closest to it). There is clear evidence for an
excess
of events in the zero multiplicity bin in the G region over that expected
from
a KNO fit (solid curve). No such excess is visible in the N region. See ref
\cite{devlin}
 for more details.}
\label{kno}
\end{figure}

CDF tags two jets of rapidity $\eta_1$  and $\eta_2$ ($\eta_2$ is assumed
to be
greater  than  $\eta_1$). They  then look at  the the   multiplicity of
charged
tracks in region G defined  by  $\eta_1<\eta_g <\eta_2$ and region N
defined by
the remainder of the  rapidity range. N then covers  the rapidity range
between
each jet and its parent (anti)proton. If color singlet exchange is
contributing
to the jet production, then one should  expect events with zero
multiplicity in
region G. Region N always has color flow across it and can therefore be
used as
a control region. A KNO type multiplicity plot is made for the G and N
regions,
see figure  \ref{kno}.
The shapes of  the distributions in  the G and N regions
are the same except for an excess in the zero multiplicity bin in the G
region.
This   provides  clear  evidence  for  events  with a  rapidity  gap  at a
rate
$f=(0.86\pm 0.12)$\%. There is no evidence for any dependence of $f$ on
either
the   transverse   momentum  of the  jets   ($E_t$)  or the   width of  the
gap
($\eta_2-\eta_1$). The rate is too  large to be due to photon exchange and
must
represent the exchange of  another color singlet  object. The obvious
candidate
is the pomeron.

These  data  leave  several   questions   unanswered.  $f$  cannot be
directly
interpreted in terms of  the strength of the  coupling of the pomeron to
quarks
and gluons since, once two jets are  produced by this mechanism, we do not
know
how often  particles are emitted into  the gap region by  the rest of the
event
and  hence  what  fraction  of  these  events  survive  to be   detected by
the
experiments.  (Bjorken\cite{bjgap}  uses the term survival  probability $S$
for
this.) Hence    $f=S\frac{pomeron-rate}{perturbative  QCD-rate}$. More data
are
needed on the $E_t$ dependence of $f$.  If, for example, the pomeron
couples to
quarks and gluons  with a form factor  as opposed to a  hard coupling, then
one
would expect $f$ to fall as $E_t$ increases. A constant $f$ would indicate
that
it coupled in a similar way to gluons.

\begin{figure}
\epsfbox{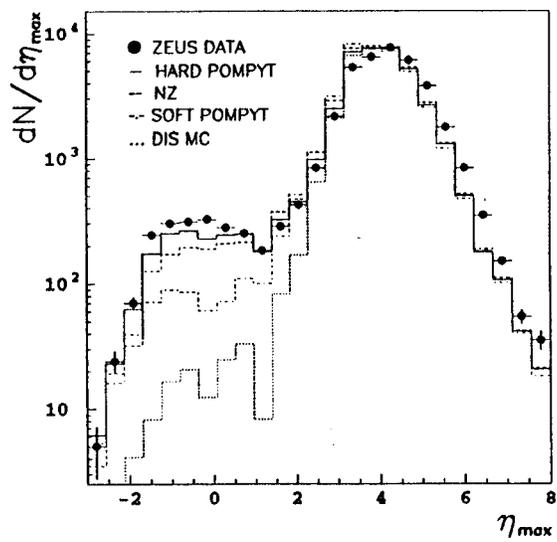}
\caption[]{The $\eta_{max}$
distribution (see text) as measured by the ZEUS collaboration
\cite{zeus}.
}
\label{zeus}
\end{figure}

A similar phenomenon has been observed at HERA. In the usual picture of
deep-inelastic scattering a quark is struck by the virtual photon and
ejected
from the target proton. This quark then hadronizes into a jet (the current
jet)
and since its color is compensated by the target remnant, particles are
produced in the rapidity region between the current jet and the beam
proton.
$\eta_{max}$ is defined as the rapidity of the particle with the largest
rapidity in a particular event. (The proton is initially moving in the
positive
rapidity direction.) One would expect that there are always particles
produced
near the initial proton and so the $\eta_{max}$ distribution would have a
peak
at large positive value. Figure \ref{zeus} shows the distribution as
measured
by ZEUS \cite{zeus}. The data show, in addition to the expected peak, a
large
number of events where $\eta_{max}$ is very small. Approximately 8\% of the
events have no hadrons in the direction of the initial proton; the fraction
is
independent of  the mass $Q^2$ of the virtual photon. Similar phenomena
have
been observed by the H1 collaboration\cite{h1gap}

The rate of events in this
region of $\eta_{max}\sim 0$ is much larger than expected from a
Monte-Carlo
based on this picture of Deep inelastic scattering. The excess of events
can be
explained if there is some color neutral component of the proton which
itself
can be disassociated by the virtual photon. This component will have some
fraction of the proton's momentum. After interaction with the virtual
photon
the hadronization and color neutralization need only take place among the
fragments of this color neutral system. There is no necessity for particle
production in the rapidity region between the object and its parent proton.
A second peak at smaller values of $\eta_{max}$ will then appear. One
candidate
for this object is the pomeron \cite{ingelmann},
which can be though of as an
object similar to other hadrons with quark constituents. A model of this
type
where the object (pomeron) has a structure function of the form
 $f(x)\sim x(1-x)$ is compatible with the ZEUS data\cite{zeus}.

\begin{figure}
\epsfbox{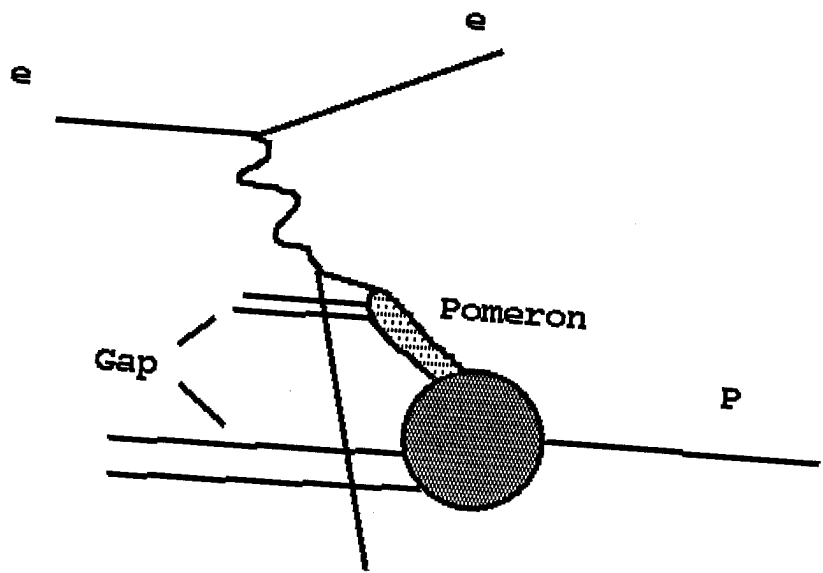}
\caption[]{The simplest contribution to deep inelastic scattering in QCD
with the
possibility to produce an event with low $\eta_{max}$. The simplest object
in
QCD to play the role of the pomeron is a two gluon system.}
\label{pomeron}
\end{figure}

The simplest candidate for this object in QCD is a two-gluon object
\cite{nussinov}  as shown in figure \ref{pomeron}. This simple picture has
been extended to and builds up the BFKL pomeron \cite{nikolaev}.  This
picture
is also in qualitative agreement with the data. Note that, as in the case
of
events with rapidity gaps at hadron colliders, the relative normalization
of
these color singlet pieces is difficult to extract from the data.

\section{Particle Multiplicity in Heavy Quark Jets.}

\begin{figure}
\epsfbox{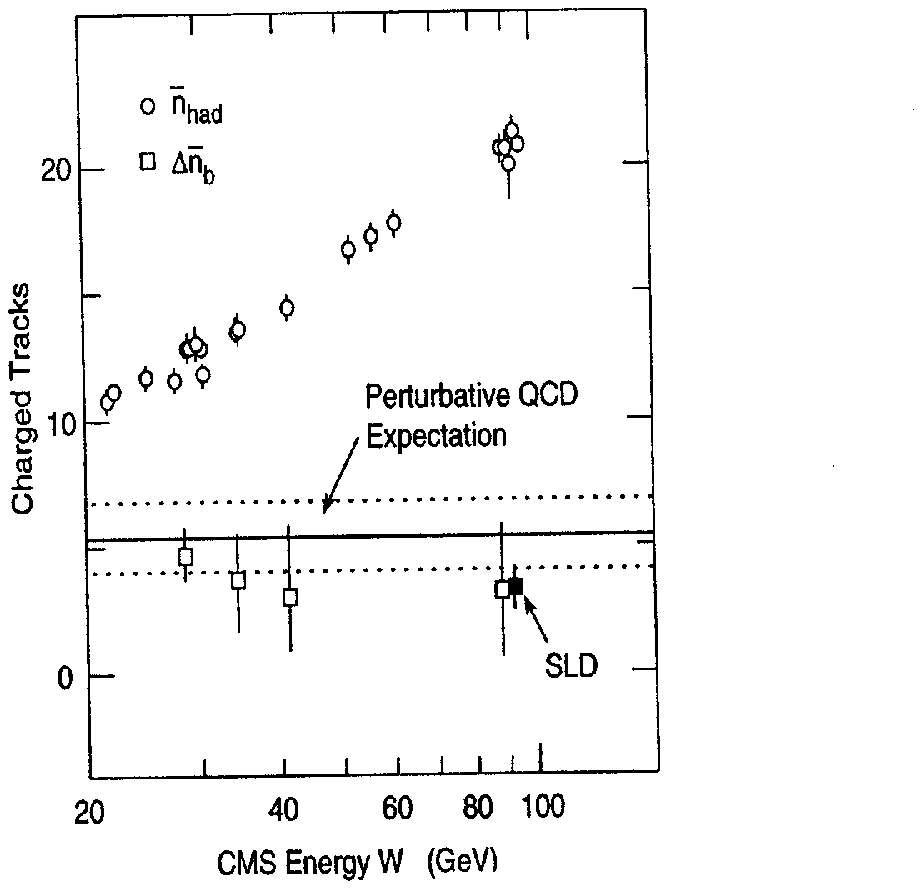}
\caption[]{The behaviour of the charged particle multiplicity in $e^+e^-$
events
as a function of $\sqrt{s}$
\cite{sldmult}.
 The plot shows the total
multiplicity $n_{had}$ as well as the difference in multiplicity between
events tagged as being from or not from the production of a $b\overline{b}$
 pair ($\Delta n_b$)}
\label{sld}
\end{figure}

While the particle  multiplicity cannot be  calculated in perturbative QCD,
its
growth  with  energy can be.  Consider  the  radiation of  a gluon  off a
quark
produced of mass $M$ say in $e^+e^-$  annihilation. If the gluon has energy
$E$
and is emitted at angle  $\theta$ with respect to  the quark, then the
emission
probability                                                     behaves
 as
$$d\sigma=\frac{\theta^2d\theta^2}{\theta^2+\delta^2}\frac{dE}{E}$$    This
has
two consequences,  radiation at $\theta<\delta$ is  suppressed resulting a
what
is called a  ``dead-cone'' and a heavy  quark radiates  less than a light
quark
\cite{khoze}.  We should  expect the particle  multiplicity  ($N$) from a
heavy
quark  pair  ($Q$)  to be  less  than  that   from  a  light  quark  pair
($q$)
$$N(Q\overline{Q},\sqrt{s})=N(q\overline{q},\sqrt{s})-N(q\overline{q},M)$$
Note
that the difference in  multiplicities is  independent of $\sqrt{s}$. The
naive
expectation             based    on    the        available      phase
space
$$N(Q\overline{Q},\sqrt{s})=N(q\overline{q},\sqrt{s}-2M)$$
predicts a
difference that is not  independent of $\sqrt{s}$.  Figure \ref{sld} shows
data
at  various      energies\cite{sldmult}.  While  the  total   charged
particle
multiplicity  rises with  $\sqrt{s}$,  the  difference between
multiplicity in
$b-$quark events (tagged using the finite $b$ lifetime) and average events
does
not.

\section{Conclusions.}

The past few years has seen a continuing development in our understanding
of
QCD. The strong coupling constant is now known to a precision of order 5\%
{}.
Its precision can only be expected to improve slowly in the near future
since
most of the measurements are now limited by various theoretical
uncertainties.
An improved measurement of the hadronic width of the $Z$ from LEP is on of
the
few areas where a more precise measurement of a physical quantity will
yield a
more accurate value of $\alpha_s$. The recent developments involving
lattice
gauge theory calculations with propagating light quarks is another area
where
one can hope for increased precision.

There are still some experimental results that, while accommodated by
perturbative QCD, are not entirely satisfactorily explained. While the long
standing problem of the prompt photon rate in $p\overline{p}$ collisions
may
now by going away, the production rate of bottom quarks is still not fully
digested. Interesting data on $\psi$ and $\Upsilon$ production from CDF are
yet to be fully understood.

Much interest, both theoretical and experimental, has occurred in the area
of
semi-hard (or small-$x$) QCD. Diffractive phenomena, for a long time
dismissed
as incalculable and hence uninteresting, are finally being given the
attention
that they deserve. There are many ``predictions'' for phenomena in this
region
of phase space. However, a systematic procedure for calculating the
subleading
corrections to the BFKL equation is lacking. Such a procedure is badly
needed,
for, until we can determine the size of these terms, we cannot say how
accurate
predictions using BFKL can be expected to be.

The preparation of this talk took place while I was a visitor in the
FERMILAB theory group. I am grateful to Keith Ellis and the other members
of
the group for their hospitality.
This work was supported by the Director, Office of Energy
Research, Office of High Energy and Nuclear Physics, Division of High
Energy Physics of the U.S. Department of Energy under Contract
DE-AC03-76SF00098.


\begin{thebibliography}{99}
\bibitem{rpp} Review of Particle Properties \pr{D50}{1174}{94}.
\bibitem{Khadra93} A.X. El-Khadra \etal,  \prl{69}{729}{92}.
A. X. El-Khadra, presented at 1993 Lattice conference, OHSTPY-HEP-T-93-020,
A.X. El-Khadra \etal, FNAL 94-091/T (1994)
\bibitem{luscher}  Martin Luscher, Rainer Sommer,Peter Weisz and
Ulli Wolff \np{B413}{481}{94}.
\bibitem{lepage} C.T.H. Davies \etal
\ OHSTPY-HEP-T-94-013, Aug 1994.
\bibitem{davies} G.P. Lepage, and
 J. Sloan presented at 1993 Lattice conference, hep-lat/9312070
\bibitem{cleo} R. Balest, \etal \  CLEO-CONF-94-28, Jul 1994.
\bibitem{lepshape} For a recent review see S. Bethke
PITHA-94-29, Aug 1994.
\bibitem{sldshape} K. Abe, \etal \prl{71}{2528}{94}.
\bibitem{ronan} D.A. Bower \etal \ LBL-35812, M. Aolki \etal \
Submitted to the XXVII International Conference on High Energy Physics
(Glasgow
1994), D. Decamp \etal \ \pl{B284}{163}{92}.
\bibitem{h1} S. Soldner-Rembold, submitted to Int. Conf. on High Energy
 Physics, Glasgow,
Scotland, Jul 20-27, 1994.
\bibitem{inglemann}  G. Ingelman,J. Rathsman, TSL-ISV-94-0096, May 1994.
\bibitem{cleogamma}CLEO collaboration, reported at the QCD94 Conference in
Montpellier.
\bibitem{alephfrag} G.Cowan,
submitted to Int. Conf. on High Energy Physics, Glasgow,
Scotland, Jul 20-27, 1994.
\bibitem{delphifrag}P. Abreu, \etal \ \pl{B311}{408}{93}.
\bibitem{braaten} {S. Narison and A. Pich \pl{B211}{183}{88},
 E. Braaten, S. Narison
and A. Pich, \np{B373}{581}{92}}
\bibitem{sumrule} {M.A. Shifman, A.I Vainshtein, and V.I. Zakharov
 \np{B147}{385}{79}.}
\bibitem{cleotau} J. Alexander, \etal CLEO-CONF-94-26, Jul
1994, Submitted to Int. Conf. on High Energy Physics, Glasgow,
Scotland, Jul 20-27, 1994.
\bibitem{alephtau} D. Buskulic,\etal \pl{B307}{209}{93}.
\bibitem{alephtaunew} ALEPH collaboration reported at the QCD94 Conference
in
Montpellier.
\bibitem{cdfb} F. Abe, \etal \prl{71}{2396}{93},
\prl{71}{500}{93} and CDF collaboration these proceedings.
\bibitem{d0b} G. Alves and T. Huen
for the D0 collaboration, these proceedings.
\bibitem{nde} P. Nason, S. Dawson and R.K. Ellis, \np{B303}{607}{88}.
\bibitem{cdftop} F. Abe, \etal \ \prl{73}{225}{94}.
\bibitem{laenen} E. Laenen J. Smith, and W.L. van Neerven,
\pl{B321}{254}{94}.
\bibitem{d0top} S. Abachi, \etal \  \prl{72}{138}{94}.
\bibitem{stirling} R. Baier and R. Ruckl, \zp{c19}{251}{83}, F. Halzen
\etal
\ \pr{D30}{700}{84}.
\bibitem{fleming} E. Braaten  and T.-C. Yuan \prl{71}{1673}{93}.
\bibitem{fleming1} E. Braaten \etal \  \pl{B333}{548}{94}.
\bibitem{roy} D.P. Roy and K. Sridhar, CERN-TH-7434-94.
\bibitem{vaia} Vaia Papadimitriou, \etal \ FERMILAB-CONF-94-221-E, Aug
1994.
\bibitem{vogelsang} M. Gluck, \etal \ DO-TH-94-02-REV, Feb 1994.
\bibitem{owens} H. Baer, J. Ohnemus and J. Owens  \pr{D42}{61}{90}
\bibitem{grv}  M. Gluck, E. Reya, and  A. Vogt \pl{B306}{391}{93}.
\bibitem{cdfgamma} F. Abe, \etal \
FERMILAB-PUB-94-208-E, 1994.
\bibitem{D0gamma} S. Fahey for the D0 collaboration, these proceedings.
\bibitem{cteq} James Botts, \etal \ \pl{B304}{159}{93}.
\bibitem{herarev} For a review see, for example, Gunter Wolf
DESY-94-178,1994.
\bibitem{GRV}M. Gluck, E. Reya and M.Voit, \etal \zp{C53}{127}{92}.
\bibitem{mrs} A.D. Martin,  R.G. Roberts and W.J. Stirling
HEPPH-9409257, Jul 1994.
\bibitem{d0gamma} D0Collaboration, these proceedings.
\bibitem{blair} R. Blair, \etal \ FERMILAB-CONF-94-269-E.
\bibitem{lhc}For a recent review see, for example, J.F. Gunion, UCD-94-24,
(1994).
\bibitem{gampair} B. Bailey, J.F. Owens and J. Ohnemus \pr{D46}{2018}{92}.
\bibitem{glap}V. N. Gribov and L.N. Lipatov \sjn{15}{438}{72},
L.N. Lipatov, \sjn{20}{181}{74},
Yu. L. Dokshitser, {\it Sov. Phys. JETP}
{\bf 46}, 641 (1977), G. Altarelli and G. Parisi \np{B126}{298}{77}.
\bibitem{bfkl}E.A. Kurayev, L.N. Lipatov and V.S. Fadin, {\it Sov. Phys.
JETP}
{\bf 45}, 119 (1977), Ya. Ya. Bailitsky and L.N. Lipatov,
\sjn{28}{882}{78}.
\bibitem{ellis}By R.K. Ellis, Z. Kunszt and E.M. Levin, \np{B420}{517}{94}.
\bibitem{lipatov} L.N. Lipatov  {\it Sov.Phys.JETP} {\bf63},904 (1986).
\bibitem{mueller1} A.H. Mueller and B. Patel \np{B425}{471}{94}, A.H.
Mueller \np{B415}{373}{94}, N.N. Nikolaev, B.G. Zakharov
and V.R. Zoller J.Exp.Theor.Phys. {\bf 78}, 806 (1994).
\bibitem{mueller}  A.H. Mueller, H. Navelet \np{B282}{87}.
\bibitem{schmdt} Vittorio Del Duca and Carl R. Schmidt, DESY-94-114.
W.J. Stirling \np{B423}{56}{94}.
\bibitem{d0rap}D0 Collaboration, these proceedings.
\bibitem{herwig}G. Marchesini and B.R. Webber, \np{B310}{461}{88}.
\bibitem{devlin}F. Abe, \etal \ FERMILAB-PUB-194-94E, {\it Phys. Rev. Lett.
}
(to appear).
\bibitem{d0gap}S. Abachi, \etal \ \prl{72}{2332}{94}.
\bibitem{bjgap} For a recent review see, J.D. Bjorken, SLAC-PUB-6463,
(1994),
\bibitem{zeus} M. Derrick, \etal \ \pl{B332}{228}{94}.
\bibitem{h1gap}T. Ahmed, \etal \ DESY-94-133 (1994).
\bibitem{ingelmann} G. Ingelmann and P.E. Shlein, \pl{152B}{256}{85}.
\bibitem{nussinov}S. Nussinov, \pr{d41}{246}{76}, S. Donnachie
 and P.V. Landshoff, \np{B244}{322}{84}.
\bibitem{nikolaev}N. Nikolaev and B.G. Zakharov,  \zp{C53}{331}92{}
\bibitem{khoze} B. A. Schumm, L. Dokshitser  and V, A. Khoze
\prl{69}{3025}{92}.
\bibitem{sldmult} K. Abe, \etal \ \prl{72}{3145}{94}.
\end{thebibliography}
\end{document}